\documentclass[12pt]{iopart}

 \usepackage{graphicx,caption,amssymb,relsize}
\begin{document}

\title{Electromagnetic Hyper-Lift: optical nano-tweezers with hyperbolic materials}

\author{Evgenii E. Narimanov}

\address{School of Electrical and Computer Engineering and Birck Nanotechnology Center, \\ Purdue University,  
West Lafayette, Indiana 47907, USA}
\ead{evgenii@purdue.edu}
\vspace{10pt}

\begin{abstract}
Optical tweezers, formed by tightly focused propagating laser beams, offer the unique capability to trap 
and control microscopic particles over a broad size range.
However, the diffraction inherent to propagating optical fields, limits the resulting  resolution and the accuracy of 
particle manipulation. Here we show that the phenomenon of ``auto-focusing'' inherent to hyperbolic materials
in cylinder geometry, can be used for spatial control with nanometer accuracy. Furthermore, due to 
highly efficient light focusing in hyperbolic media that is not restricted by diffraction, the resulting electromagnetically induced 
forces exceed those of conventional optical tweezers by several orders of magnitude, which allows more efficient
particle manipulation at reduced illumination intensity.  
\end{abstract}

\section{Introduction}

Optical tweezers \cite{McGloin2006,Ashkin1970,Ashkin1986} are highly focused laser beams used to trap and manipulate microscopic particles, from biological cells \cite{Ashkin1987}  to  individual molecules.\cite{Ruttley2024}  Pioneered by Arthur Ashkin in 1970 \cite{Ashkin1970}  and developed in the 1980s,\cite{Ashkin1986} optical tweezers have now become invaluable tools in many fields, from  molecular biology \cite{Choudhary2019} and biophysics\cite{Matthew2024} to nanotechnology,\cite{Sneh2024}  enabling researchers to study the mechanical properties of DNA,\cite{Peterman2023} measure forces within cells,\cite{Arbore2019} and investigate molecular interactions at the nanoscale.\cite{Eechampati2022}

It is however very difficult to directly trap nanometer-sized targets with conventional optical tweezers.\cite{Choudhary2019} 
In order to achieve a stable trapping,  the  potential generated by a laser beam must be much higher than the thermal fluctuations of the particle in the medium. As the polarizability of a dielectric object scales with its volume, for a nanometer-sized particle the trapping energy is too small to
prevent the thermal escape from the conventional optical trap. The necessary enhancement of the trapping can be achieved in direct proximity of 
plasmonic resonators, which support a rapidly varying field close to the subwavelength hot spots induced by the plasmon resonance, \cite{Novotny1997}  
leading to the development of the plasmonic optical tweezers.\cite{Grigorenko2008,Juan2011,Shoji2014} The idea of  using the rapid variation of the evanescent near-fields to optically trap nanoscale objects well below the diffraction limit was also implemented in optical tweezers based on slot waveguides \cite{Yang2009} and photonic crystals.\cite{Mandal2010}

However, while strong evanescent fields near sharp features of plasmonic and dielectric nanoresonators \cite{Kim2008} allow to reliably trap nanometer-size targets,\cite{Choudhary2019}  these are essentially stationary optical traps that do not offer the key capability of conventional optical tweezers -- the ability to move objects in space by manipulating the spatial variation of the optical beams forming the trap. Here, we introduce the alternative approach to optical nano-tweezing, based on subwavelength light focusing and confinement enabled by hyperbolic optical materials. Subwavelength ``hot spot'' that originate from the 
``hyper-lensing''  of light in the hyperbolic medium,\cite{OE2006,EnghetaPRB2006,hyper-lithography} are not confined to the proximity of sharp features in the medium (such as e.g. a ``corner'' of a dielectric resonator), but are defined by the corresponding  dielectric permittivity tensor and its frequency dependence. As a result, with a continuous variation of the illumination frequency, the hyperbolic ``hot spots'' can be  moved in space \cite{HSI}  -- thus offering the capability for controlled spatial manipulation of nanoparticles at a deeply subwavelength scale.     

\begin{figure}[htbp] %  figure placement: here, top, bottom, or page
\begin{center}
\small
 \includegraphics[width=5.5in]{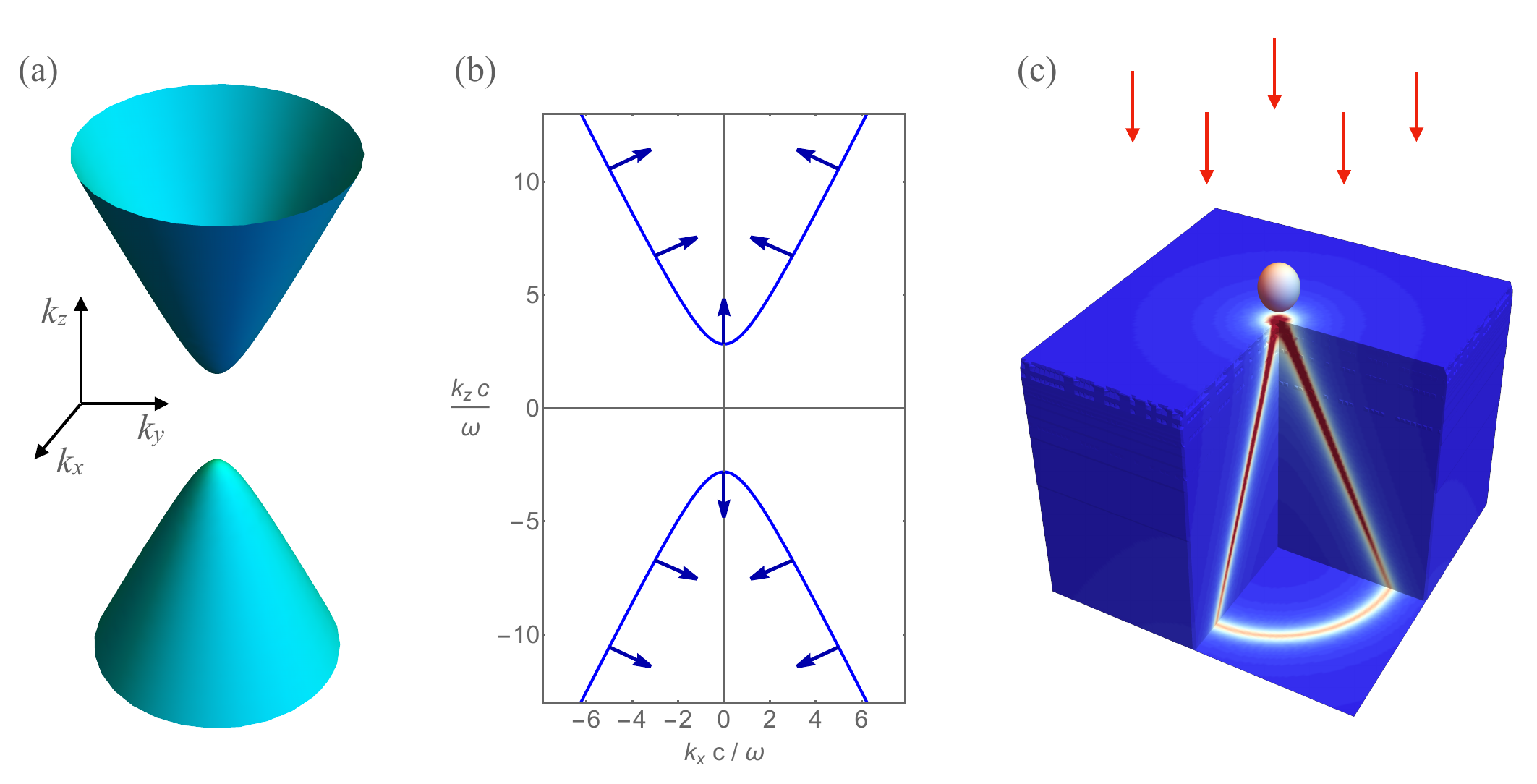} 
   \caption{Light propagation and scattering in a hyperbolic medium.
   Panel (a) : the iso-frequency surface for light in a uxiaxial hyperbolic material 
   with ${\rm Re}\left[\epsilon_z\right] < 0$, ${\rm Re}\left[\epsilon_x\right]\equiv {\rm Re}\left[\epsilon_x\right] > 0$ (corresponding to e.g. lower-frequency hyperbolic band of hexagonal boron nitride ($h$BN) \cite{hBN1,hBN2}  -- see Fig. \ref{fig:eps}(a)).  
   Panel (b) :  The cross-section of the iso-frequency surface in (a) by the plane $k_y = 0$. The arrows normal to the iso-frequency curve, indicate the direction of the group velocity of light. Note that, aside from the limited wavenumber interval of $\left| k_x c/\omega\right| \lesssim 1$,  waves
with different  wavevectors  propagate at essentially the same angle with respect to the crystal symmetry axis $\hat{\bf z}$, forming the  hyperbolic ``light cone''.
   Panel (c) : The cut-out view of  the hyperbolic medium with light incident on a subwavelength  (nano)object (particle radius $R \ll \lambda_0$) at its  top surface. The nanoparticle scatters light into the medium with in-plane numbers $k \gg \omega/c$, forming the hyperbolic light cone -- with intensity shown here in false color.   }
   \normalsize
   \label{fig:cone}
   \end{center}
 \end{figure}

\section{Light in Hyperbolic Media}

In a uniaxial medium, TM-polarized propagating waves are characterized by the dispersion
\begin{eqnarray}
{\epsilon_\parallel} {k_\parallel^2} + {\epsilon_\perp} {k_\perp^2}& = &{\epsilon_\perp} {\epsilon_\parallel} \ {\omega^2} / {c^2},
\label{eq:dispersion}
\end{eqnarray}
where the subscripts $\parallel$ and $\perp$ represent the directions respectively parallel and perpendicular to the symmetry axis. When the real parts of the  two orthogonal components of the permittivity, $\epsilon_\parallel$ and $\epsilon_\perp$, have opposite sign, the corresponding iso-frequency
surface is a hyperboloid (see Fig. \ref{fig:cone}(a) ), formed by either one (${\rm Re}\left[ \epsilon_\parallel \right] > 0$,  ${\rm Re}\left[ \epsilon_\perp \right]  <  0$) or two sheets (${\rm Re}\left[ \epsilon_\parallel \right] > 0$,  ${\rm Re}\left[ \epsilon_\perp \right]  <  0$).  In the limit of large wavenumbers
$k \gg \omega/c$, the hyperbolic iso-frequency surface asymptotically approaches the cone
\begin{eqnarray}
 {k_\parallel^2} & = &  - \frac{\epsilon_\perp}{\epsilon_\parallel}  {k_\perp^2}.
\end{eqnarray}
As the group velocity (and the Poynting vector) in the low-loss limit are orthogonal to the iso-frequency surface,  for $k \gg \omega/c$ the electromagnetic waves in the uniaxial hyperbolic medium  propagate at the angle 
\begin{eqnarray}
\vartheta & = & \arctan \left[ {\rm Re} \ \sqrt{- \epsilon_\perp / \epsilon_\parallel} \ \right]
\label{eq:theta}
\end{eqnarray}
with respect to the symmetry axis $\hat{\bf z}$ (see Fig. \ref{fig:cone}(b)). As a result, light scattered into the hyperbolic medium from a nanoparticle at its surface, forms the  conical sheet pattern -- see Fig. \ref{fig:cone}(c).

For the electric field of the light scattered into the hyperbolic material, at the distance $r$ much larger than the  separation $h$ between 
the nanoparticle and the material surface, but much smaller than the free space wavelength  $\lambda_0$, $h \ll r \ll \lambda_0$, we obtain
\begin{eqnarray}
{\bf E} & = & - {\bf \nabla}\left({\bf p \cdot\nabla}  
\frac{1}{\sqrt{{\epsilon_\perp}{\epsilon_\parallel} \rho^2 +{\epsilon_\perp^2} z^2}}
\right) ,
\end{eqnarray}
where ${\bf p}$ is the effective dipole moment of the nanoparticle (that includes the contributions of the particle polarization induced 
by the incident field, as well as that of the resulting surface charge), $z$ and $\rho$  correspond to the distance from the emitter in the directions respectively parallel and orthogonal to the symmetry axis of the material. 

For a lossless hyperbolic medium,  in  the emission cone defined by Eqn (\ref{eq:theta}), $ \rho  \to  \pm z \tan\vartheta$, 
 the scattered electric field is singular: $E \to \infty$. This singularity is however removed by any finite amount of loss, with the electric field in the hyperbolic emission cone sheet 
\begin{eqnarray}
E\left(   \rho =   z  \tan\vartheta\right) & \propto & \frac{p}{r^3} \ Q^{5/2}, 
\label{eq:cone-field}
\end{eqnarray}
where the hyperbolic quality factor $Q$ is defined as 
\begin{eqnarray}
Q & \equiv & 
\left| \frac{{\rm Re} \ \sqrt{-  \epsilon_\perp / \epsilon_\parallel } }{{\rm Im} \ \sqrt{-   \epsilon_\perp  / \epsilon_\parallel}} \right|.
\end{eqnarray}

\begin{figure}[htbp] %  figure placement: here, top, bottom, or page
\begin{center}
\small
 \includegraphics[width=6.in]{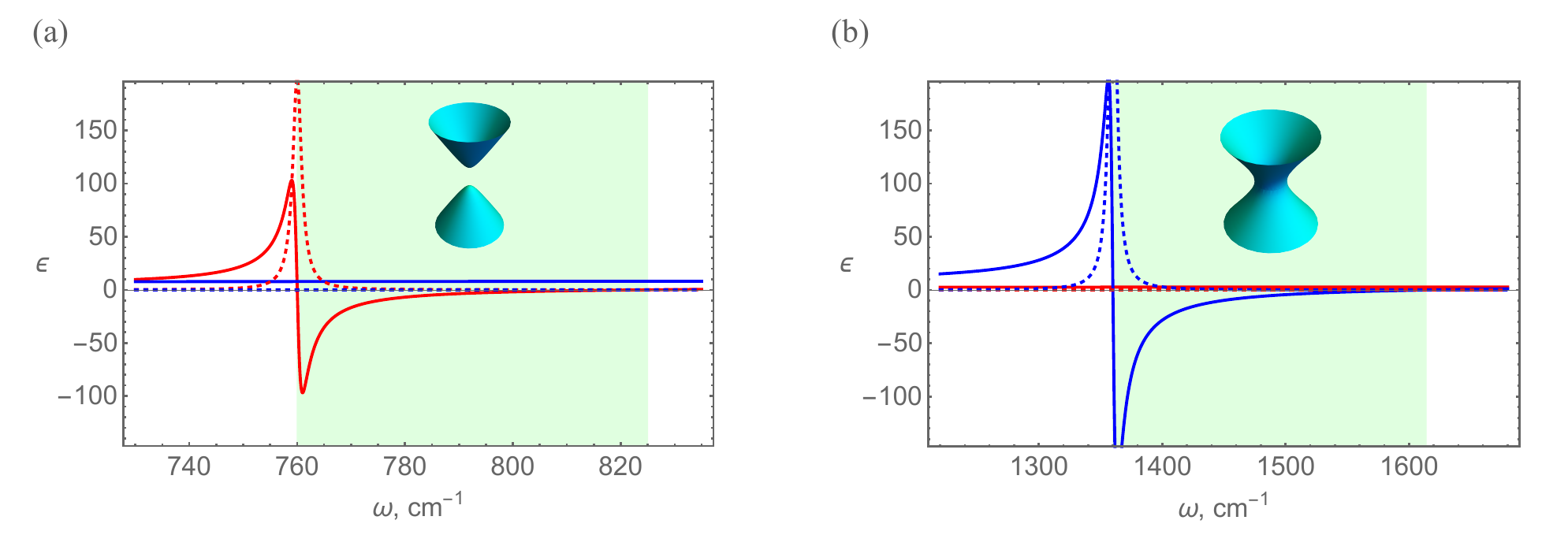} 
   \caption{The dielectric permittivity of the hexagonal boron nitride ($h$BN), showing its (a) lower- and (b)  higher-frequency hyperbolic bands.\cite{hBN1,hBN2}  The red and blue colors show respectively the dielectric permittivity components parallel ($\epsilon_\parallel$) and perpendicular ($\epsilon_\perp$) to the symmetry axis of $h$BN. Solid lines correspond to real and dashed curves to the imaginary parts of the permittivity. Light green shading shows the regions with the hyperbolic response, and insets visualize the corresponding topology of the iso-frequency surface for the waves propagating in the material in these bands. 
   }
   \normalsize
   \label{fig:eps}
   \end{center}
 \end{figure}

For natural hyperbolic materials such as the hexagonal boron nitride \cite{hBN1,hBN2,hBN3,hBN4,hBN5} the quality factor close to the center of the hyperbolic band $Q \gtrsim 30$, reaching the value of   $Q \simeq 100$ in isotopically purified high-quality crystals.\cite{hBNisotope} So while the limiting value of (\ref{eq:cone-field}) is finite, in the hyperbolic emission cone the scattered field intensity $I \sim E^2$   is enhanced by {\it four orders of magnitude}. 
 
This dramatic field enhancement in natural hyperbolic media  is comparable to the ultimate performance limit of plasmonic nanostructures.\cite{Lu2024} However, while in  plasmonic systems the location of ``hot spot'' of the enhanced optical field is  defined by the geometry of the nano-resonator, the position of the hyperbolic emission cone
\begin{eqnarray}
\rho  & = & z  \cdot \tan\vartheta\left(\omega\right) \equiv z \cdot  {\rm Re}  \ \sqrt{-
 \epsilon_\perp\left(\omega\right)  / \epsilon_\parallel\left(\omega\right)  } 
\end{eqnarray}
is directly controlled by the illumination frequency $\omega$. Furthermore, due to the relatively narrow hyperbolic bands in the frequency response of polar dielectric crystals such as $h$BN, the hyperbolic emission cone completes the ``full sweep'' of the material while 
$\vartheta$ changes from $0$ to $\pi/2$ over the illumination wavelength variation by $ \sim 10\%$ (e.g. from $\omega \simeq  760 \ {\rm cm}^{-1}$ to $\omega \simeq  825 \ {\rm cm}^{-1}$ in the lower frequency hyperbolic band and from $\omega \simeq  1360 \ {\rm cm}^{-1}$ to $\omega \simeq 1610 \ {\rm cm}^{-1}$ in the higher frequency hyperbolic band of $h$BN -- see Fig. \ref{fig:eps}).

The unique combination of strong subwavelength optical intensity localization and nearly unlimited spatial control of the electromagnetic field by the variation of the illumination wavelength makes natural hyperbolic materials an ideal platform for the optical nano-tweezers. 

\section{Auto-Focusing in Hyperbolic Resonators}

\begin{figure}[htbp] %  figure placement: here, top, bottom, or page
\begin{center}
\small
 \includegraphics[width=6.in]{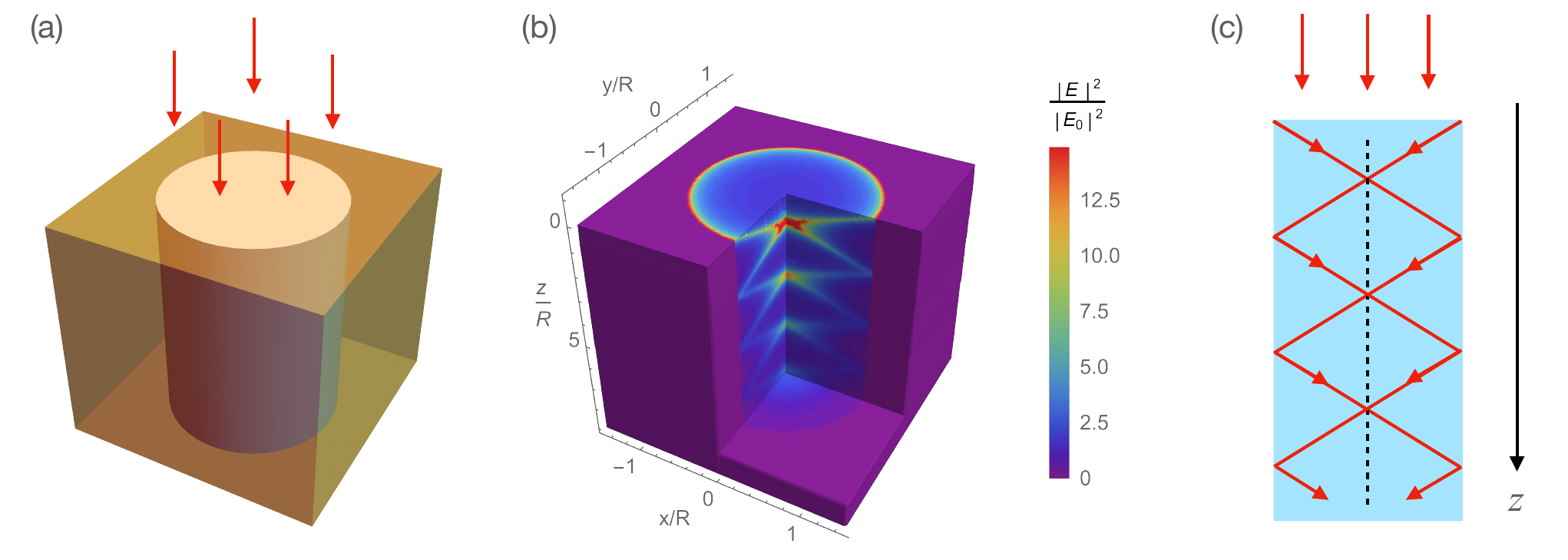} 
   \caption{Auto-focusing in a hyperbolic cylinder resonator. 
   Panel (a) : the schematics of the system geometry. Light (red arrows) is incident on a cylinder formed by a uniaxial hyperbolic material
   with its symmetry axis in the longitudinal direction. Gold metal cladding, surrounding the cylinder, is not necessary  and only used here 
   to suppress the intensity outside the resonator for easier visualization of the electric field inside the hyperbolic cylinder in (b).  
   Panel (b) : the cut-out view of the hyperbolic cylinder resonator with the electric field intensity shown in false color.
   Panel (c) : the origin of the auto-focusing in the hyperbolic cylinder, illustrated in the cross-section of the resonator (shown by the light-blue shading). The incident light (vertical red arrows) is scattered into the hyperbolic medium by the sharp edge of the resonator. Due to the essential feature of wave propagation at an almost constant angle with respect to the symmetry  axis of the hyperbolic material, the scattered light forms the converging conical sheet pattern (diagonal red arrows), leading to the first focus at the axis of the cylinder. Further light propagation now yields a diverging conical sheet pattern, until specular reflection at the vertical interface of the cylinder turns in back onto to convergence to the next focus. The process is then repeated multiple times. 
   }
   \normalsize
   \label{fig:multi-foci}
   \end{center}
 \end{figure}

The unique conical sheet pattern on light propagation in a uniaxial hyperbolic medium,  lead to the phenomenon of optical ``auto-focusing''
in a cylindrical waveguide with subwavelength cross-section. With uniform illumination (see Fig. \ref{fig:multi-foci}(a)) the rotational symmetry of the problem together with the requirement of scattering at the angle $\vartheta$ with respect to the symmetry axis, implies the converging
conical pattern of light scattered into the cylinder by diffraction at its edge -- see Fig. \ref{fig:multi-foci}(a),(b).

Once converging to first (``primary'') focus  at the symmetry axis, the conical pattern will expand until specular reflection at the cylinder boundary, followed by the repetition of this process {\it ad invinum}, resulting in the formation of the multiple focal spots at
\begin{eqnarray}
z_n = \left(2 n -1\right) z_f\left(\omega\right)  \equiv   \left(2 n -1\right) \cdot R \cdot \cot\vartheta\left(\omega\right),
\end{eqnarray}
where $R \ll \lambda_0$ is the cylinder radius, and $n = 1,\ldots$ is an integer. The formation of the primary ($n=1$) auto-focus has been 
observed in the experiments of Ref. \cite{He2021}, albeit in the ``reverse'' -- with the point  source positioned at the auto-focus,  leading to the light emission into the far field due to scattering from the cylinder edge.

Note that, most of the light scattered into the subwavelength hyperbolic cylinder by its sharp edge, is characterized by large (in comparison with $\omega/c$) wavenumbers that can only be supported in the hyperbolic medium, so that  all multiple reflections  at the vertical wall of the cylinder ( see Fig. \ref{fig:multi-foci}(c)) are total. 

\begin{figure}[htbp] %  figure placement: here, top, bottom, or page
\begin{center}
\small
 \includegraphics[width=6.in]{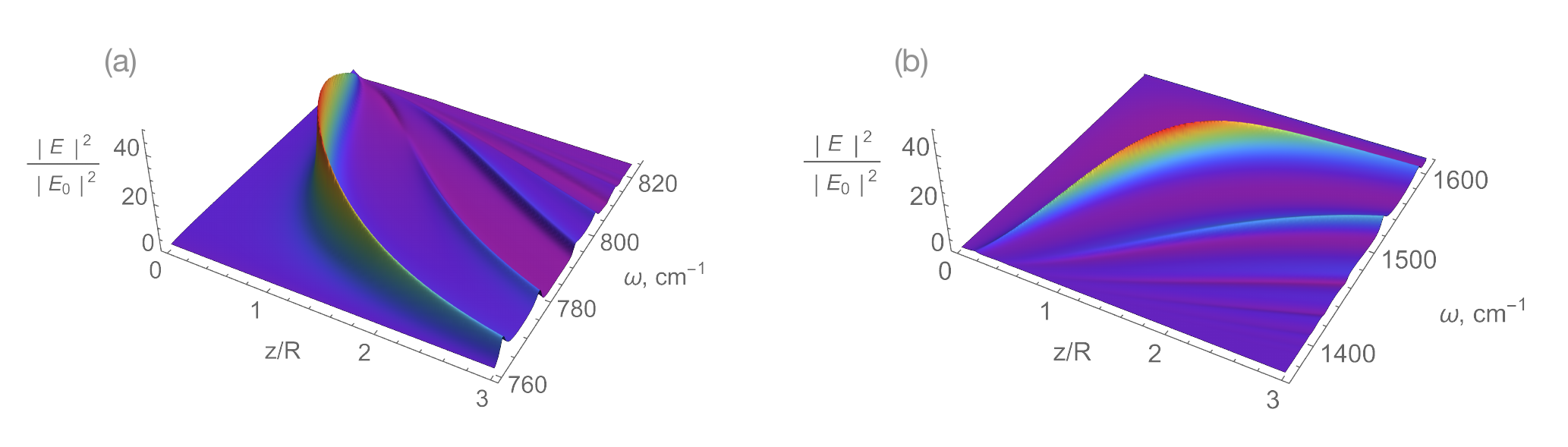} 
   \caption{The electric field intensity (normalized to the intensity of the incident light) on the axis of the $h$BN cylinder resonator in its low (a) and (b) high frequency hyperbolic bands 
   }
   \normalsize
   \label{fig:map}
   \end{center}
 \end{figure}

However, the material absorption in the hyperbolic medium leads to the progressive weakening of the higher-order foci $z_n$. The key issue here is not the overall field amplitude reduction, as the latter can always be compensated by the increase of the illumination intensity up to the material damage threshold (that is notoriously high in the hyperbolic $h$BN \cite{Ginsberg2023,Tancogne2018,Chen2024} 
$\sim 50 \ {\rm TW}/{\rm cm}^2$) -- but the spacial broadening of the focal spots with the increase of propagation distance in the hyperbolic medium. In the cylinder geometry, the absorption coefficient for a high-$k$ wave propagating in a hyperbolic material is proportional to its wavenumber,
\begin{eqnarray}
{\rm Im} \left[ k_z \right] \propto \frac{{\rm Re} \left[ k_z \right]}{Q},
\end{eqnarray}
so that the higher-$k$ components are absorbed at the faster rate -- leading to the noticeable broadening of the high-order ``hot sports''.
This effect is clearly visible in Fig. \ref{fig:map} where we show the field intensity at the axis of an $h$BN cylinder (normalized to that of the incident light) as a function of the distance from the top interface $z$ and and the illumination frequency $\omega$, for both its (a)  lower-  and (b) higher-frequency hyperbolic bands.

For the electric field on the axis of the cylinder we obtain
\begin{eqnarray}
\frac{E}{E_0} & = & 1 + i \sqrt{2} \ \frac{\epsilon_\parallel - \epsilon_0}{\epsilon_0 + i \cdot \sqrt{- \epsilon_\parallel \epsilon_\perp}}
\frac{\epsilon_\perp}{\sqrt{{\epsilon_0^2} - {\epsilon_\parallel \epsilon_\perp}}} 
\nonumber \\\ & \times & 
\exp\left[
- i  \ {\rm sign}\left[ {\rm Re} \ \epsilon_\perp\right] \ 
\sqrt{- \frac{\epsilon_\perp}{\epsilon_\parallel}} \ 
\left( \frac{3 \pi}{4} - \arctan\left( \frac{\epsilon_0}{\sqrt{-{\epsilon_\parallel}{\epsilon_\perp}}} \right) \right) \frac{z}{R} \right]
\nonumber \\ & \times & 
{\rm Li}_{1/2}\left[ \exp\left(  i \pi \ {\rm sign}\left[ {\rm Re} \   \epsilon_\perp\right]  \left( \sqrt{- \frac{\epsilon_\perp}{\epsilon_\parallel}} \frac{z}{R} - 1\right) \right) \right],
\label{eq:E1}
\end{eqnarray} 
where $\epsilon_0$ is dielectric permittivity of the medium surrounding the hyperbolic cylinder, and $Li_{1/2}$  is the polylogarithm function of order $1/2$.
For a relatively small loss ($Q \gg 1$), close to the focal spot of the $n$-th order, 
\begin{eqnarray}
z = z_n + \delta z, \ \ \ \left| \delta z\right| \ll z_f,
\end{eqnarray}
Eqn. (\ref{eq:E1}) can be reduced to 
\begin{eqnarray}
\frac{E}{E_0} & = & 1 + i \sqrt{2} \ \frac{\epsilon_\parallel - \epsilon_0}{\epsilon_0 + i \cdot \sqrt{- \epsilon_\parallel \epsilon_\perp}}
\frac{\epsilon_\perp}{\sqrt{{\epsilon_0^2} - {\epsilon_\parallel \epsilon_\perp}}} 
\nonumber \\\ & \times & 
\exp\left[
- \ i  \ {\rm sign}\left[ {\rm Re} \ \epsilon_\perp\right] \ 
\left( 2 n - 1 \right) \ 
\left( \frac{3 \pi}{4} - {\rm Re}  \arctan\left( \frac{\epsilon_0}{\sqrt{-{\epsilon_\parallel}{\epsilon_\perp}}} \right) \right) \right]
\nonumber \\ & \times & 
\frac{1}{
\sqrt{
\mathlarger{\frac{2 n - 1}{Q}  }
 + i   \ {\rm sign}\left[ {\rm Re} \ \epsilon_\perp\right]  \mathlarger{ \frac{\delta z}{z_f} }
 }
}.
\label{eq:E2}
\end{eqnarray} 

From Eqn. (\ref{eq:E2}), the width of the intensity hot spot is increasing linearly with the order the focus,
\begin{eqnarray}
\Delta z_n & = & z_f \ \frac{2 n - 1}{Q}, 
\label{eq:dzf}
\end{eqnarray}  
and it is the first order focal spot that yields to strongest optical confinement. For a cylinder with the radius of $R = 100$ nm, that was fabricated from a natural (i.e. not isotopically purified) $h$BN,  for the frequencies close to center of the hyperbolic bands we find the size of the primary focal spot 
$\Delta z_1 \simeq 10$ nm.

\begin{figure}[htbp] %  figure placement: here, top, bottom, or page
\begin{center}
\small
 \includegraphics[width=6.in]{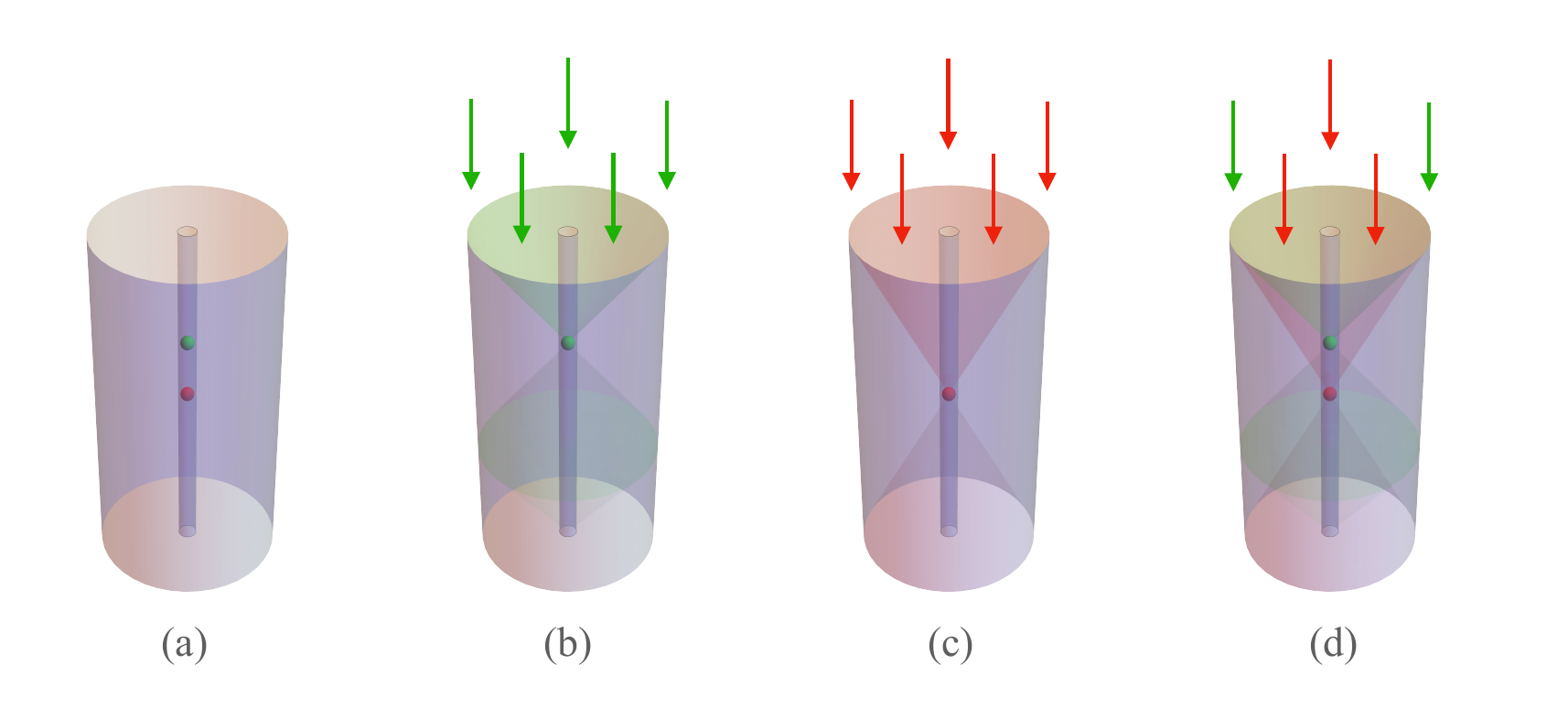} 
   \caption{
   Panel (a) : the schematics of the optical ``hyper-lift'', formed by an open channel at the symmetry axis of a hyperbolic material cylinder.
   Panels (b), (c) :  light scattered by the edge of a hyperbolic material cylinder is naturally focused at the cylinder axis;  
   the position of the focus is defined by the ratio of the different components of the dielectric permittivity of the hyperbolic medium, 
   and due to the inherent material dispersion strongly depends on the illumination wavelength.  Panel (d): simultaneous illumination 
   by several different wavelengths yields different foci, which allows independent manipulation of multiple particles by continuous variation 
   of the optical frequencies.
   }
   \normalsize
   \label{fig:hyper-lift}
   \end{center}
 \end{figure}

\section{Optical ``Hyper-Lift'' and Hyperbolic Nano-tweezers}

\begin{figure}[htbp] %  figure placement: here, top, bottom, or page
 \centering
 \includegraphics[width=5.25in]{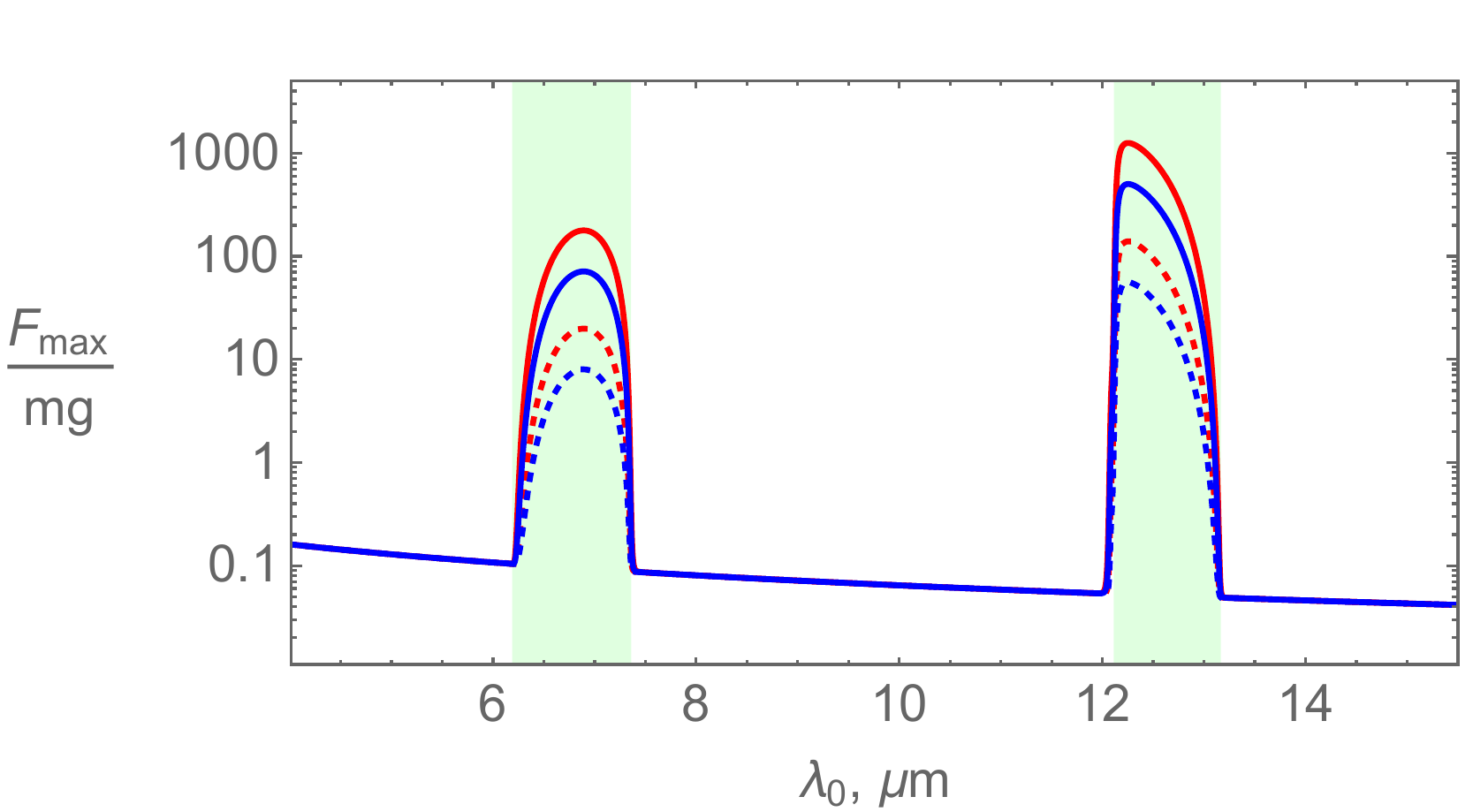} 
 \caption{The optical force acting on a spherical nano-particle in the cylinder ``tunnel' at the axis of hexagonal boron nitride cylinder (see Fig. \ref{fig:hyper-lift}), normalized  to the particle weight, as a function of optical illumination wavelength $\lambda_0$. Light green shading shows the wavelength intervals with hyperbolic response in $h$BN.\cite{hBN1,hBN2}  The cylinder radius is $R = 200 \ {\rm nm}$ (red lines) and 
 $R =  500 \ {\rm nm}$ (blue lines). The nanoparticle permittivity $\epsilon_p = 1.96$ and its mass density $\rho_p = 2.2 \ {\rm g}/{\rm cm}^3$ (corresponding to fused silica),  its radius  Solid and dashed curves show  the trapping force at the primary ($n = 1$) and secondary ($n=2$) foci respectively. 
The incident field amplitude $E_0 =  200$ V/cm, with the intensity $I_0 \simeq 50 \ {\rm W/cm}^2$ (corresponding to a focused laser pointer).}
   \label{fig:Fmg}
 \end{figure}

High optical intensity at the axis of the hyperbolic material cylinder and tight subwavelength focusing, can be used for highly efficient trapping of the small particle  with nanometer-scale spatial control and manipulation. Introducing a narrow open channel at the axis of the hyperbolic cylinder 
(see Fig. \ref{fig:hyper-lift}) with the radius smaller than the size of the hot spot (\ref{eq:dzf}), 
\begin{eqnarray}
r_c < \Delta z_1 \equiv z_f/Q,
\end{eqnarray}
we find the electric field in the channel accurately described by Eqn. (\ref{eq:E2}),  since the radial field component  that is discontinuous at the channel surface, vanishes in the limit  $\rho \to 0$. Then, even for incident light intensity $I \sim 50 \  {\rm W}/{\rm cm}^2$, corresponding to the 
output strength of a laser pointer, the  optical forces acting on a nanoparticle  in the narrow channel of a subwavelength  hyperbolic cylinder, 
substantially exceed gravity -- see Fig. \ref{fig:Fmg} , thus allowing for optical levitation in such ``hyper-lift''. For the ratio of the (maximum) optical force acting on a trapped particle $F_{\rm max}$, to its weight $mg$, in a hyperbolic band we obtain
\begin{eqnarray}
\frac{F_{\rm max}}{mg} & = &  
\frac{1}{2 \sqrt{3} \pi} \  {\rm Re \sqrt{-\frac{ \epsilon_\perp }{ \epsilon_\parallel}} } \ \left| \frac{ \epsilon_\perp\left( \epsilon_\parallel - \epsilon_0\right)  }{ \epsilon_0^2 -\epsilon_\parallel \epsilon_\perp}\right|^2 \cdot \frac{\epsilon_p - \epsilon_0}{\epsilon_p + 2 \epsilon_0} \cdot \frac{E_0^2}{\rho_p g R} \cdot Q^2, 
\end{eqnarray}
where $E_0$ is the amplitude of the optical field incident on the cylinder, $\epsilon_p$ and $\rho$ are respectively the dielectric permittivity and the material density of the trapped particle, and $\epsilon_0$ is the permittivity of the medium surrounding the cylinder and filling in the channel.

The strength and efficiency of the nanoparticle trapping in optical tweezers is characterized by the optical trap stiffness $\alpha$ defined from
\begin{eqnarray}
F & = & - \kappa \cdot \delta z, 
\end{eqnarray}
where $F$ is the restoring force that an optical trap  exerts on a nanoparticle when it is displaced from the trap center by the distance $\delta z$.
For trapping at the primary focus ($n=1$) we  obtain
\begin{eqnarray}
\kappa & = &\kappa_0 \cdot \frac{3 }{4 \pi^2} \  {\rm Re \sqrt{-\frac{ \epsilon_\perp }{ \epsilon_\parallel}} }
\left|
\frac{\epsilon_\perp\left( \epsilon_\parallel - \epsilon_0\right) }{\epsilon_0^2 - \epsilon_\parallel \epsilon_\perp  }
\right|^2
\cdot Q^3 
\cdot
 \frac{\lambda_0^2 }{R^2},
\end{eqnarray}
where $\kappa_0$ is the optical trap stiffness of the conventional (free space) optical tweezers where the maximum light intensity $I_0$ is equal to that of the field incident on the ``hyper-lift''. For a subwavelength hyperbolic cylinder, $R < \lambda_0/10$ while for existing low-loss natural hyperbolic materials (such as e.g., $h$BN, sapphire, etc. \cite{naturally-hyperbolic}) $Q > 10$. As a result, the hyper-lift positioned in the field  of conventional optical tweezers, will increase the resulting optical trap stiffness by {\it more}  than {\it five orders of magnitude}. 

For a stable optical trapping, the potential energy of the nanoparticle in the optical field,
\begin{eqnarray}
U & = & \frac{\alpha \ \langle E^2\rangle }{2},
\end{eqnarray}
where $\langle E^2\rangle $ is the (time-averaged) intensity and $\alpha$ is the polarizability of the trapped object, must exceed the thermal energy $k_B T$,
\begin{eqnarray}
U \gg k_B T.
\label{eq:UkT}
\end{eqnarray}
In general,
\begin{eqnarray}
\alpha =  \frac{3}{4 \pi} \ c_0 \ v_p,
\end{eqnarray}
where $v_p$ is the particle volume, and the dimensionless factor $c_0$ depends on the shape and composition of the object, e.g. for a spherical 
nanoparticle with the dielectric permittivity $\epsilon_p$
\begin{eqnarray}
c_0 & = & \frac{\epsilon_p - \epsilon_0}{\epsilon_p + 2 \epsilon_0}.
\end{eqnarray}
For the illumination intensity $I_0 \equiv c E_0^2 / 8 \pi$ needed for stable trapping (\ref{eq:UkT}) in the primary ($n=1$) focus of the central channel of the optical hyper-lift we therefore obtain
\begin{eqnarray}
I_0 \gg I_c \equiv \frac{c}{8 \pi}  \left| \frac{ \epsilon_0^2 - \epsilon_\parallel\epsilon_\perp }{\epsilon_\perp \left( \epsilon_\parallel - \epsilon_0 \right)} \right|^2 
\cdot   \frac{1}{Q} \frac{k_B T}{v}
\simeq \frac{c}{8 \pi} \frac{k_B T}{Q v}.
\label{eq:I_0}
\end{eqnarray} 
The square of the hyperbolic quality factor in the denominator of (\ref{eq:I_0}) substantially lowers the incident intensity needed to trap individual atoms and molecules at room temperature, as for a high quality natural hyperbolic materials such as isotopically purified $h$BN, $Q  \sim 100$. E.g. for a medium-size atom, with $v_p \simeq 10^{-23}$ cm$^3$, we find the  critical intensity $I_c \simeq 5 \ {\rm GW}/{\rm cm}^2$, which is four orders of magnitude lower than  the material damage threshold in the hexagonal boron nitride.\cite{Ginsberg2023,Tancogne2018}

Furthermore, due to the  strong material dispersion inherent to hyperbolic media, the use of several illumination fields at different frequencies offers the capability of parallel trapping of several particles, while tuning these individual frequencies allows controlled spatial movement of the trapped objects -- see Fig. \ref{fig:hyper-lift}(d).

 \section{Discussion}

Highly efficient light focusing in hyperbolic media, unlimited by optical diffraction, offers the optical trapping capabilities not readily 
available in other platforms. As the signal-to-noise ratio in experiments with optical tweezers,\cite{Neuman2004}  typically increases linearly with the optical trap stiffness, its improvement  by five orders of magnitude in the hyperbolic tweezers is a major advantage of the proposed approach.

Furthermore, the potential to simultaneously and independently control the spatial locations of several individual atoms in the channel of the proposed hyper-lift (see Fig. \ref{fig:hyper-lift}(d)) leads to the intriguing possibility of the ``optical reactor''  -- where a chemical reaction is ``driven'' in real time by optical beams from several wavelength-tunable mid-IR lasers.

 \section{Acknowledgements}

This work was partially supported by the National Science Foundation DMREF program, Award 2119157.

\end{document}